\documentclass[prl,aps,preprint,showpacs,superscriptaddress,floatfix]{revtex4}
\usepackage[english]{babel}
\usepackage{color}
\usepackage{graphicx}
\usepackage{amsmath}
\usepackage{amsfonts}
\usepackage{amssymb}
\usepackage[squaren]{SIunits}
\usepackage{color}
\usepackage[colorlinks=true,linkcolor=blue,citecolor=blue,pdfauthor=Gross]{hyperref}

\begin{document}

%\preprint{T. Helm {\it et al.}, version: 13 August 2010}

\title{Magnetic Breakdown in the electron-doped cuprate superconductor 
Nd$_{2-x}$Ce$_x$CuO$_4$: the reconstructed Fermi surface survives in the 
strongly overdoped regime}

\author{T. Helm}
\affiliation{Walther-Mei{\ss}ner-Institut, Bayerische Akademie der Wissenschaften,
Walther-Mei{\ss}ner-Str. 8, D-85748 Garching, Germany}

\author{M. V. Kartsovnik}
\email{Mark.Kartsovnik@wmi.badw-muenchen.de}
\affiliation{Walther-Mei{\ss}ner-Institut, Bayerische Akademie der Wissenschaften,
Walther-Mei{\ss}ner-Str. 8, D-85748 Garching, Germany}

\author{I. Sheikin}
\affiliation{Laboratoire National des Champs Magn\'{e}tiques Intenses, CNRS 25 rue
des Martyrs B.P. 166, 38042 Grenoble Cedex 9, France}

\author{M. Bartkowiak}
\affiliation{Hochfeld-Magnetlabor Dresden, Forschungszentrum
Dresden-Rossendorf, Bautzner Landstr. 400, D-01328 Dresden, Germany}

\author{F. Wolff-Fabris}
\affiliation{Hochfeld-Magnetlabor Dresden, Forschungszentrum
Dresden-Rossendorf, Bautzner Landstr. 400, D-01328 Dresden, Germany}

\author{N. Bittner}
\affiliation{Walther-Mei{\ss}ner-Institut, Bayerische Akademie der Wissenschaften,
Walther-Mei{\ss}ner-Str. 8, D-85748 Garching, Germany}

\author{W. Biberacher}
\affiliation{Walther-Mei{\ss}ner-Institut, Bayerische Akademie der Wissenschaften,
Walther-Mei{\ss}ner-Str. 8, D-85748 Garching, Germany}

\author{M. Lambacher}
\affiliation{Walther-Mei{\ss}ner-Institut, Bayerische Akademie der Wissenschaften,
Walther-Mei{\ss}ner-Str. 8, D-85748 Garching, Germany}

\author{A. Erb}
\affiliation{Walther-Mei{\ss}ner-Institut, Bayerische Akademie der Wissenschaften,
Walther-Mei{\ss}ner-Str. 8, D-85748 Garching, Germany}

\author{J. Wosnitza}
\affiliation{Hochfeld-Magnetlabor Dresden, Forschungszentrum
Dresden-Rossendorf, Bautzner Landstr. 400, D-01328 Dresden, Germany}

\author{R. Gross}
\affiliation{Walther-Mei{\ss}ner-Institut, Bayerische Akademie der Wissenschaften,
Walther-Mei{\ss}ner-Str. 8, D-85748 Garching, Germany}
\affiliation{Physik-Department, Technische Universit\"{a}t M\"{u}nchen, James Franck
Stra{\ss}e, D-85748 Garching, Germany}

\begin{abstract}
We report on semiclassical angle-dependent magnetoresistance oscillations
(AMRO) and the Shubnikov-de Haas effect in the electron-overdoped
cuprate superconductor Nd$_{2-x}$Ce$_x$CuO$_4$. Our data provide
convincing evidence for magnetic breakdown in the system. This shows that a
reconstructed multiply-connected Fermi surface persists, at least at strong
magnetic fields, up to the highest doping level of the superconducting regime.
\end{abstract}

\date{\today}

\maketitle

The recent discovery of magnetic quantum oscillations in the resistivity of
hole-~\cite{doir07,yell08,bang08,vign08} and electron-doped~\cite{helm09}
cuprate superconductors clearly demonstrates the importance of high-field
magnetotransport experiments for exploring the Fermi surface in these
materials. In particular, quantum oscillations of the magnetoresistance,
the Shubnikov-de Haas (SdH) effect, observed in the electron-doped
cuprate Nd$_{2-x}$Ce$_x$CuO$_4$ (NCCO) have proved to be very useful
for tracing changes in the Fermi surface with varying carrier
concentration~\cite{helm09}. For
$x=0.17$, corresponding to nearly the border of the superconducting regime on
the overdoped side, high-frequency SdH oscillations were found which are in
perfect agreement with the large cylindrical Fermi surface predicted by band
structure calculations~\cite{mass89} and angle-resolved photoemission
spectroscopy (ARPES)~\cite{armi02,mats07}. However, lowering the Ce
concentration by just 1\% leads to their suppression and the emergence of much
slower oscillations~\cite{helm09}. The slow oscillations observed for
optimally doped ($x = 0.15$) and slightly overdoped ($x = 0.16$) NCCO crystals
indicate that the Fermi surface is reconstructed in this doping range, at least
at high magnetic fields. It is, thus, tempting to suggest a quantum
phase transition associated with a translational symmetry breaking at a doping
level between $x = 0.16$ and $0.17$. This is supported by in-plane
magnetotransport data at lower magnetic fields on another electron-doped
superconductor Pr$_{2-x}$Ce$_x$CuO$_4$~\cite{daga04} implying a quantum
critical point at $x \approx 0.165$. However, a similar experiment on
NCCO~\cite{lamb08} rather indicates two small groups of carriers and,
hence, a reconstructed Fermi surface persisting up to $x = 0.17$. Moreover, it
was pointed out \cite{helm09,chak10}, the observation of fast SdH oscillations does not
unambiguously rule out the Fermi surface reconstruction for $x =
0.17$: if the potential responsible for the symmetry breaking is nonzero but
weak, such oscillations can arise as a result of magnetic breakdown. Thus, the
question of whether the quantum phase transition occurs inside the
superconducting doping range or the reconstructed Fermi surface survives
throughout the entire superconducting dome is still open and one of the key
issues in high temperature superconductivity.

Another important question is related to the exact shape of the reconstructed
Fermi surface in overdoped NCCO. It was proposed~\cite{helm09,chak10} that the
reconstruction occurs due to a $(\pi/a,\pi/a)$ density wave ordering
($a$ is the lattice constant in the CuO$_2$ plane). Indeed, the measured
frequency of the slow SdH oscillations is consistent with the size of small
hole pockets, which should be formed around $(\pi/2a,\pi/2a)$ due to such
ordering. However, to verify this scenario it is desirable to
determine not only the size but also the shape of the Fermi pockets. Towards
this end, a very efficient method which has been widely used for mapping the
in-plane Fermi surfaces of organic conductors~\cite{kart04} and other layered
systems such as Sr$_2$RuO$_4$~\cite{berg03} and intercalated
graphite~\cite{baxe98}, is semiclassical angle-dependent magnetoresistance
oscillations (AMRO). This is a geometric effect directly related
to the shape of a weakly warped cylindrical Fermi
surface~\cite{yama89,kart92,huss03,grig10}. It has also been observed
in hole-overdoped Tl$_2$Ba$_2$CuO$_{6+\delta}$ (Tl2201)~\cite{huss03,huss06}.
and used to extract the shape of the Fermi surface both
within the plane of the conducting layers and in the out-of-plane direction, as
well as to evaluate the scattering anisotropy. However, Tl2201 is
the only member of the cuprate family in which AMRO have been found so far.

Here, we report on the interlayer resistivity of overdoped NCCO studied as a
function of the orientation and strength of the applied magnetic field. We
present angle-dependent magnetoresistance patterns for crystals with $x = 0.16$
and $0.17$ which exhibit distinct features characteristic of AMRO. Most
remarkably, for both doping levels the AMRO positions are found to be very
similar. Together with the earlier SdH data~\cite{helm09}, this result implies
magnetic breakdown in the system. This conclusion is
confirmed by a new high-resolution SdH experiment on strongly overdoped ($x =
0.17$) NCCO which clearly reveals the presence of a weak superlattice potential
up to the highest doping level.

The NCCO crystals used in our experiments were grown by traveling solvent
floating zone technique and characterized as described elsewhere~\cite{lamb10}.
For the measurements of the angle-dependent magnetoresistance, the samples were
mounted on a two-axes rotating stage allowing {\it in-situ} rotations at low
temperatures, at a fixed magnetic field up to 28 T generated by the 20 MW
resistive magnet at the Grenoble High Magnetic Field Laboratory. The interlayer
resistance was measured as a function of polar angle $\theta$ between the field
direction and crystallographic [001] axis of the sample, at different fixed
azimuthal angles $\varphi$, as shown in Fig. 1(a). The SdH experiment was done
at the Hochfeld-Magnetlabor Dresden in pulsed fields up to 65 T
applied perpendicular to CuO$_2$ layers.

\begin{figure}[tb]
\centering{
\includegraphics[width=0.7\columnwidth]{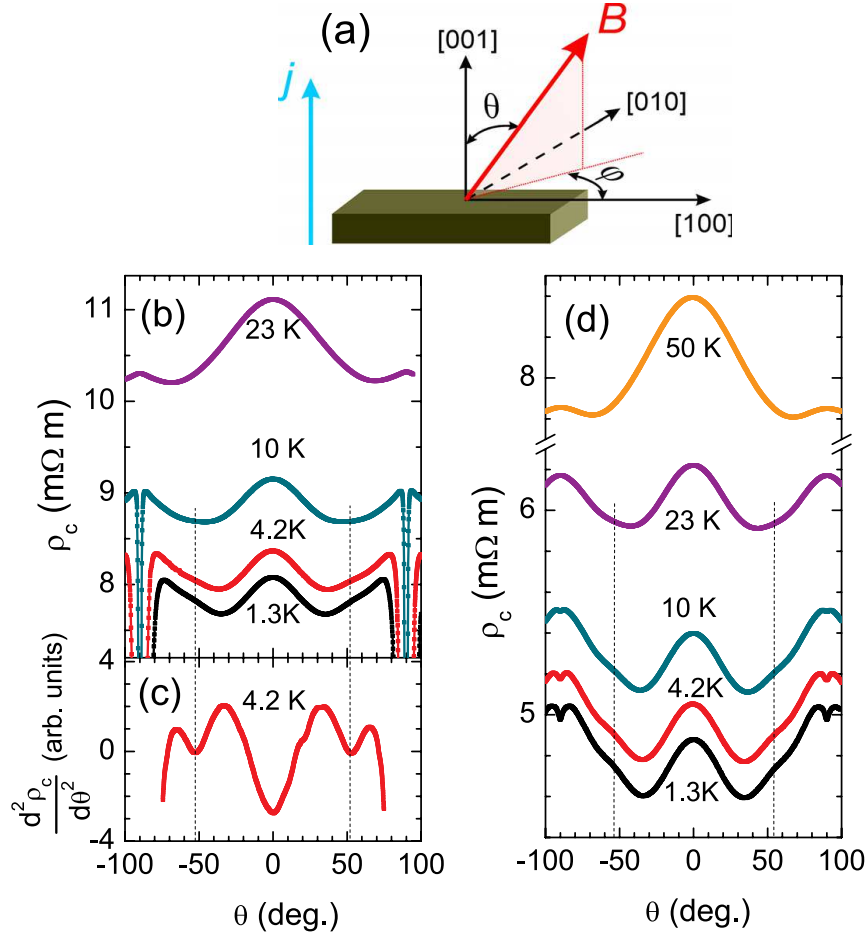}}
\caption{(Color online)(a) Sketch of the experimental configuration.
(b) Interlayer resistivity $\rho_c$ of a slightly overdoped NCCO
sample ($x=0.16$) as a function of tilt angle $\theta$, at different
temperatures. The range $|\theta|<75^{\circ}$ corresponds to the normal
conducting state; the rapid drop of $\rho_c$ outside this range is due
to onset of superconductivity. The vertical dashed lines are drawn through
the AMRO features centered at $|\theta|\approx 53^{\circ}$, which are independent
of temperature and magnetic field strength. (c) Second derivative of the
resistivity in (b) exhibiting dips at the AMRO peak positions.
(d) Same as (b), for a strongly overdoped sample ($x=0.17$).
}
\label{AMRO}
\end{figure}

Fig. \ref{AMRO}(b),(d) shows the angle-dependent interlayer resistivity
$\rho_{\rm c}$ of NCCO with $x = 0.16$ and $0.17$, respectively, recorded at
different temperatures for $\varphi = 45^{\circ}$ and $B = 28$\,T. At this
field, superconductivity at low temperatures is manifest by a sharp dip
within a narrow angular range around $\pm 90^{\circ}$.
In the normal state, we observe a prominent maximum at $\theta =
0^{\circ}$ and, in addition, a pair of shallow humps superimposed on the
monotonic slope around $\theta \approx \pm 53^{\circ}$. To illustrate them more
clearly, Fig.~\ref{AMRO}(c) shows the second derivative $d^2\rho_{\rm
c}/d\theta^2$ exhibiting dips at these angles. These features, though rather
weak, were reproduced for several samples at both doping levels.
To the accuracy of the experiment, the angular positions of the features
are independent of temperature, see Fig.~\ref{AMRO}.
We have also checked that they do not shift on reducing the field strength
down to 23\,T. This is exactly what one expects for AMRO, whose positions
are solely determined by the Fermi surface geometry.

For $T\leq 10$ K, the magnetoresistance behavior shown in Fig.~\ref{AMRO}
resembles that observed earlier on hole-overdoped Tl2201~\cite{huss03,huss06}.
For the latter compound, it was shown that both the central hump and the side
features originated from the AMRO effect on the large cylindrical Fermi
surface. In particular, the resistivity peak at $\theta =0^{\circ}$ was
considered as a fingerprint of a warped Fermi cylinder centered at the corner
of the Brillouin zone and satisfying the symmetry requirements of a
body-centered tetragonal lattice~\cite{huss03}. Since NCCO has the same crystal
symmetry, it is tempting to also attribute the central hump in our
$\rho_{\rm c}(\theta)$ curves to the large Fermi cylinder. 
We note, however, that unlike at $\theta\approx \pm 53^{\circ}$, 
the field direction corresponding to $\theta = 0^{\circ}$ coincides 
with the crystal symmetry axis normal to the layers, which should always 
lead to an extremum in the $\rho_{\rm c}(\theta)$ dependence. Therefore, 
one should not disregard other possible mechanisms, besides AMRO, which 
could cause a maximum at this field direction. Indeed, the 
experimentally observed evolution of the central hump with temperature
is opposite to that expected for AMRO. Since AMRO are an
effect of the cyclotron motion of charge carriers in a strong magnetic field,
they require a sufficiently large scattering time $\tau$. As $\tau$
increases with increasing $T$, the AMRO are expected to gradually vanish. This
is, indeed, the case for the hump-like features around $\theta \approx \pm 53^{\circ}$
in Fig.~\ref{AMRO} which can hardly be resolved above 20\,K. By contrast,
the central hump notably increases in magnitude, dominating the angular
dependence at elevated temperatures. 
In fact, the overall shape of the 23\,K curve 
for $x = 0.16$ and 50\,K curve for $x =0.17$ in Fig.~\ref{AMRO}, exhibiting 
a global maximum at ${\mathbf B}$ normal to the layers and parallel to the current 
direction, is at odds with the usual orbital effect of a magnetic field on the 
coherent interlayer charge transport. While the exact mechanism responsible 
for this anomalous behavior has still to be established, it is clear that, 
at least at high $T$, the central hump cannot be attributed to the AMRO. 
As the scattering time increases at cooling down below 20 K, the conventional 
orbital effect becomes significant, which is evidenced by the development of 
the positive slope $d\rho_{\rm c}/d|\theta|$ 
at $|\theta| \geq 30^{\circ}$ and the AMRO features at 
$|\theta| \approx 53^{\circ}$. It is also possible that the central hump is 
partly due to the AMRO at low temperatures. However, further studies 
are needed to clarify this.

\begin{figure}[tb]
\centering{
\includegraphics[width=0.5\columnwidth]{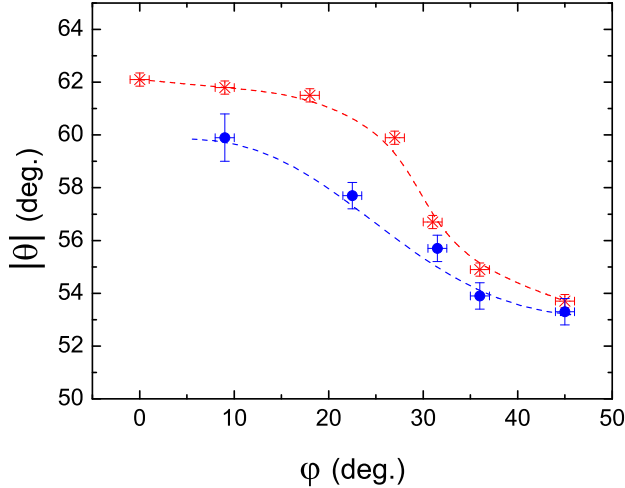}}
\caption{(Color online) Dependence of the positions of the AMRO features in
NCCO with $x=0.16$ (blue circles) and $0.17$ (red stars) on the
azimuthal orientation $\varphi$ of the field rotation plane.
Dashed lines are guides to the eye.
}
\label{AMRO_positions}
\end{figure}

The weakness of the AMRO features indicates that even at 28\,T the high field
criterion, $\omega_{\rm c}\tau \gg 1$ ($\omega_{\rm c}$ is the characteristic
cyclotron frequency), is by far not fulfilled. In this situation, simple
Yamaji-like conditions for the AMRO positions~\cite{kart04,yama89,kart92} are
not applicable and a detailed quantitative analysis of the angle-dependent
magnetoresistance is required for the determination of the relevant Fermi
surface geometry~\cite{huss03,grig10,huss06}. This is, however,
problematic in the present case due to the mentioned above coexistence of an 
anomalous and the conventional orbital magnetotransport mechanisms. 
Further experiments in an extended field and
temperature range, at different doping levels are required for separating 
these two contributions. Although
a detailed quantitative analysis is out of reach, a very important qualitative
information can be gained from the present data. In Fig.~\ref{AMRO}(b),(d)
the positions of the AMRO features around $\pm 53^{\circ}$ are
remarkably similar for $x=0.16$ and $0.17$. Moreover, as illustrated in
Fig.~\ref{AMRO_positions} they exhibit the same dependence on the azimuthal
angle $\varphi$ of the field direction. This means that the cyclotron orbits
responsible for the AMRO are identical for these two doping levels. At first
glance, this conclusion contradicts the SdH data~\cite{helm09} which exhibit a
dramatic change in the oscillation spectrum when moving from $x=0.16$ to
$0.17$. The apparent discrepancy can be resolved by taking into account the
possibility of magnetic breakdown between the hole- and electron-like parts of
the reconstructed Fermi surface.

The characteristic breakdown field $B_0$ can be estimated using the Blount
criterion~\cite{shoe84}, $\hbar\omega_0 \sim \Delta^2/\varepsilon_{\rm F}$,
where $\omega_0 = eB_0/m_{\rm c}$, $e$ is the elementary charge, $m_{\rm c}$
the cyclotron mass corresponding to the orbit on the original large Fermi
cylinder, $\varepsilon_F$ the Fermi energy, and $\Delta$ the energy gap between
the hole- and electron-like bands determined by the density-wave potential.
Using $\Delta \simeq 30$~meV~\cite{helm09}, $m_{\rm c} \approx 2\times
10^{-30}$\,kg (estimated from the $T$-dependence of the amplitude of the fast
SdH oscillations in the $x = 0.17$ sample) and $\varepsilon_F\simeq 0.5$\,eV,
we obtain $B_0 \sim 35$\,T. Although this is a rather rough estimate, it
clearly shows that magnetic breakdown can indeed be expected for overdoped NCCO
crystals within the experimentally available field range.

\begin{figure}[tb]
\centering{
\includegraphics[width=0.5\columnwidth]{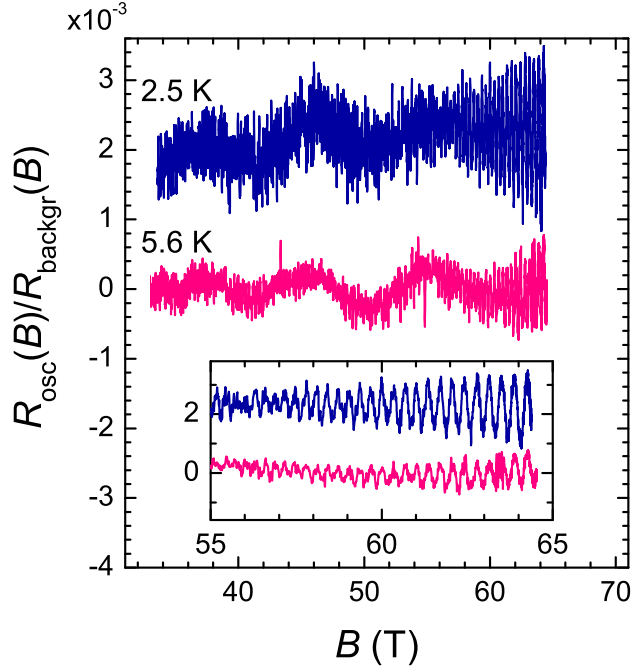}
\caption{(Color online) Slow and fast SdH oscillations in the interlayer resistance
of a strongly overdoped, $x=0.17$, NCCO crystal recorded at
two different temperatures. The data is normalized to the field-dependent
nonoscillating background resistance. The inset shows a high-field fragment of
the 2.5 K curve, demonstrating regular fast oscillations.}
}
\label{SdH}
\end{figure}

If the magnetic breakdown scenario applies to our samples, the cyclotron orbit
topology becomes dependent on the magnetic field strength. In particular, for
$B\sim B_0$ both the small classical orbits on the reconstructed Fermi surface
and the large orbit, indistinguishable from that on the original Fermi
surface, can be realized. To check this, we have carried out refined
measurements of the SdH oscillations in strongly overdoped NCCO in pulsed
magnetic fields. Fig.~\ref{SdH} shows examples of the SdH oscillations in the
interlayer resistance of an $x = 0.17$ crystal at $T \approx 2.5$ and
5.6\,K~\cite{comment-T}. Two oscillation frequencies are readily resolved at
both temperatures. Fast oscillations with frequency
$F_{\mathrm{fast}}=10.93\pm 0.05$ kT, consistent with our previous report
\cite{helm09}, are observed starting from $\approx 55$ T.
They are superimposed by slow oscillations which
can be traced down to below 40\,T. Their frequency, $F_{\mathrm{slow}}=250\pm
10$\,T, is similar to the values obtained for crystals with $x=0.15$ and
$0.16$ and attributed to small hole pockets of the reconstructed Fermi surface.

\begin{figure}[tb]
\centering{
\includegraphics[width=.5\columnwidth]{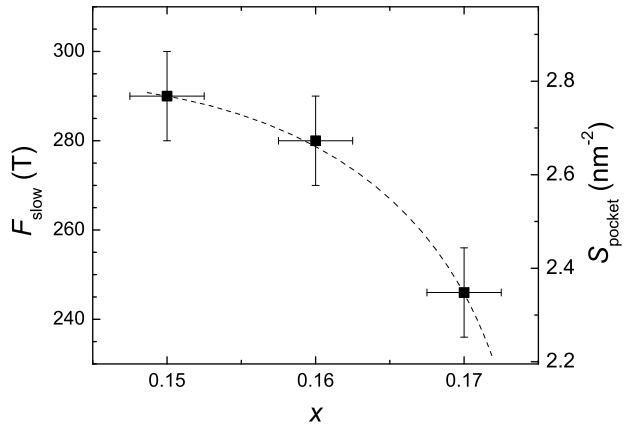}}
\caption{Frequency of the slow SdH oscillations (left scale) and the
corresponding Fermi surface area (right scale) as a function of
the electron doping level $x$. The dashed line is a guide to the eye.
}
\label{FvsX}
\end{figure}

One has to beware of sample inhomogeneity as a
possible source of coexisting SdH frequencies. Evidently, a minor volume
fraction with $x\approx 0.16$ might produce weak slow oscillations in
addition to the fast ones originating from the $x=0.17$ matrix. Therefore, we
did a careful sample characterization including X-ray and magnetization
studies, assuring a high crystal quality and doping homogeneity within $\pm
0.25\%$. However, the strongest argument regarding the origin of the slow
oscillations is provided by the doping dependence of their frequency shown
by Fig.~\ref{FvsX}. Were the oscillations caused by a small volume fraction
with a lower doping level, a change of the nominal Ce concentration $x$
would affect their amplitude but not the frequency. By contrast,
the experiment shows a clear decrease of $F_{\mathrm{slow}}$ with increasing
$x$. This is fully consistent with the expected decrease in the size
of the small hole pockets of the reconstructed Fermi surface at increasing
the electron doping.

Thus, we conclude that the slow SdH oscillations are intrinsic to NCCO with
$x=0.17$. They clearly indicate that the Fermi surface is still reconstructed
even at this doping level, the highest attainable in a bulk NCCO
crystal~\cite{lamb10}. The corresponding superlattice potential is however, 
very weak, allowing for magnetic breakdown as evidenced by the observation of the 
fast SdH oscillations. 

As noted earlier~\cite{helm09}, the presence of a superstructure in the
electronic system of overdoped NCCO seems to contradict neutron
scattering~\cite{moto07} and ARPES~\cite{armi02,mats07} results. One possibility
to resolve this apparent disagreement is to consider a hidden $d$-density wave 
ordering \cite{chak01}. On the other hand, it may be that the ordering exists 
at high magnetic fields, i.e., at the conditions of our present experiment. 
While there are some evidences supporting this scenario\cite{mats03,soni03}, 
the question is far from being settled \cite{armi10}. To this end, it would be 
highly desirable to perform magnetic spectroscopy experiments on high-quality 
overdoped NCCO crystals showing magnetic quantum oscillations.

\begin{acknowledgments}
This work was supported by the German Research Foundation via the Research Unit
FOR 538 and grant GR~1132/15, as well as by EuroMagNET II under the EC contract
228043. We also acknowledge support by the German Excellence Initiative via
NIM.
\end{acknowledgments}

%\bibliographystyle{apsrev}
%\bibliography{HTSC}

\begin{thebibliography}{25}
\bibitem[{com()}]{comment-T}
\bibinfo{note}{Overheating of the sample due to the applied transport current,
  $I = 5 -10$~mA, and fast field sweep rate was not negligible. Therefore, the
  sample temperature was estimated by comparing the superconducting transition
  in the $R(B)$ curves with those recorded at small currents, $I=10-100$
  $\mu$A, in a steady magnetic field}.
\end{thebibliography}

\end{document}